\documentclass[12pt]{elsarticle}
\usepackage{amsmath,amsfonts,amssymb,graphicx,hyperref,textcomp,color}

\DeclareMathOperator*{\SumInt}{%
\mathchoice%
  {\ooalign{$\displaystyle\sum$\cr\hidewidth$\displaystyle\int$\hidewidth\cr}}
  {\ooalign{\raisebox{.14\height}{\scalebox{.7}{$\textstyle\sum$}}\cr\hidewidth$\textstyle\int$\hidewidth\cr}}
  {\ooalign{\raisebox{.2\height}{\scalebox{.6}{$\scriptstyle\sum$}}\cr$\scriptstyle\int$\cr}}
  {\ooalign{\raisebox{.2\height}{\scalebox{.6}{$\scriptstyle\sum$}}\cr$\scriptstyle\int$\cr}}
}

\journal{Commun Nonlinear Sci Numer Simulat}

\begin{document}

\begin{frontmatter}

\title{Green's functions and the Cauchy problem of the Burgers hierarchy and forced Burgers equation}

\author[rvt]{Mathew Zuparic}
\author[rvr]{Keeley Hoek}
\address[rvt]{Defence Science and Technology Group, Canberra, ACT 2600, Australia}
\address[rvr]{Australian National University, Canberra, ACT, 2601, Australia}

\begin{abstract}
We consider the Cauchy problem for the Burgers hierarchy with general time dependent coefficients. The closed form for the Green's function of the corresponding linear equation of arbitrary order $N$ is shown to be a sum of generalised hypergeometric functions. For suitably damped initial conditions we plot the time dependence of the Cauchy problem over a range of $N$ values. For $N=1$, we introduce a spatial forcing term. Using connections between the associated second order linear Schr\"{o}dinger and Fokker-Planck equations, we give closed form expressions for the corresponding Green's functions of the \textit{sinked} Bessel process with constant drift. We then apply the Green's function to give time dependent profiles for the corresponding forced Burgers Cauchy problem. 
\end{abstract}


\end{frontmatter}


\section{Introduction}
\label{intro}

\subsection{The Burgers equation: applications and mathematical properties}
The conservative Burgers equation,
\begin{eqnarray}
\frac{\partial}{\partial t} U = \frac{\partial}{\partial z}\left( \kappa\frac{\partial}{\partial z}U - \frac{1}{2}U^2\right), \;\; \kappa \in \mathbb{R}_+, \label{BURGERS}
\end{eqnarray}
for \textit{wave-density} $U\equiv U(z,t)$, was first introduced by Burgers \cite{BURGERS1} as a simplification of the Navier-Stokes equations of fluid/gas dynamics. Since its initial appearance, Eq.(\ref{BURGERS}) and its various generalisations have been applied in a plethora of ways in the physical sciences. A non-exhaustive list includes such applications as: a canonical model for dynamic liquids and gases combining diffusion with nonlinear advection \cite{SACHDEV1}, unsaturated flow in biological systems \cite{BROADBRIDGE97}, population dynamics and biological invasion \cite{PETROVSKII05}, acoustics \cite{FOKAS05, GURBATOV12}, and computational modelling of quantum computers \cite{Yepez02, Yepez06}.

The reason behind the prolific application of Eq.(\ref{BURGERS}) since its introduction more than half a century ago is due to the seminal works of Cole \cite{COLE1} and Hopf \cite{ HOPF1}. The famous \textit{Cole-Hopf} transformation linearises the Burgers equation into the heat equation. Early studies \cite{Jeffrey1, Kuznetsov1} pointed to a globally stable solution of Eq.(\ref{BURGERS}) for a large class of initial conditions. Indeed, further fluid dynamical studies \cite{Tatsumi1, Mizushima1} found that the only way to introduce turbulence into the solution was to randomise the initial conditions. 

Addressing its mathematical properties, Taflin \cite{TAFLIN1} derived the corresponding infinite non-local conservation laws by defining Eq.(\ref{BURGERS}) using a Hamiltonian formalism. Weiss \textit{et al.} \cite{Weiss1, Weiss2} gave the \textit{Painlev\'{e}-property}, \textit{B\"{a}cklund-transform} and \textit{Lax-pair}, firmly cementing Eq.(\ref{BURGERS}) as a proto-typical integrable nonlinear partial-differential-equation (NLPDE). 

\subsection{The Burgers hierarchy}
Additionally, Olver \cite{OLVER1} showed that Eq.(\ref{BURGERS}) possesses a countably infinite number of symmetries, giving a systematic way of generating infinitely many flows which preserve Eq.(\ref{BURGERS}), namely the \textit{Burgers hierarchy}:
\begin{eqnarray}
\frac{\partial}{\partial t} U_N =  \frac{\partial}{\partial z} \left( \kappa_N \frac{\partial}{\partial z} - \frac{1}{2}U_N  \right)^N U_N, \;\; N \in \mathbb{N}. \label{hierarch}
\end{eqnarray}
The hierarchy itself has been the focus of many studies, with \cite{ABBASSBANDY10, HE13} deriving solutions to the Sharma-Tasso-Olver equation ($N=2$ instance of Eq.(\ref{hierarch})). More generally, Cao and Xu \cite{CAO10} gave a B\"{a}cklund transform for each member of the hierarchy, Adler \cite{ADLER15} gave some intriguing combinatorial properties, and Kudryashov and Sinelshchikov \cite{KUDRYASHOV1,KUDRYASHOV14} offered specific solutions, and a method to solve the Cauchy problem of Eq.(\ref{hierarch}) for arbitrary $N$.

\subsection{The forced Burgers equation}
The forced variant of the Burgers equation,
\begin{eqnarray}
\frac{\partial}{\partial t} U_{(F)} = \frac{\partial}{\partial z}\left( \kappa\frac{\partial}{\partial z}U_{(F)} - \frac{1}{2}U^2_{(F)} +2 \kappa^2 V(z,t) \right), \;\; \kappa \in \mathbb{R}_+, \label{FBURGERS}
\end{eqnarray}
with potential term $V(z,t)$, has historically received similar treatment to the conservative Burgers equation, due to the fact that Eq.(\ref{FBURGERS}) is entirely integrable, despite the introduction of the forcing term \cite{PASMANTER86}. In fact, the Cole-Hopf transform linearises Eq.(\ref{FBURGERS}) to the (non-complex) Schr\"{o}dinger equation with potential $V$. Following Calogero \cite{CALOGERO91}, the fact that  Eq.(\ref{FBURGERS}) is immediately linearisable through a global transformation \textit{and} retains its integrable qualities is in complete contrast with NLPDEs whose solution involves the \textit{Inverse Scattering Transform}. For such equations, the introduction of a forcing term generally destroys their integrability, even though they can, more often than not, be expressed in \textit{Hirota bilinear form} \cite{HIROTA76}. 

Numerical studies in \cite{JENG1, OKAMURA1} found that solutions to Eq.(\ref{FBURGERS}) with various choices of forcing always appeared to be absolutely stable, lacking the turbulent properties witnessed in other fluid dynamics inspired NLPDEs. Additionally, Kida and Sugihara \cite{KIDA2} found that the only way to introduce turbulence into the solution of Eq.(\ref{FBURGERS}) was to make the external forcing random. Kardar, Parisi and Zhang \cite{PARISI86} proposed that Eq.(\ref{FBURGERS}) with a Gaussian white noise forcing term --- the KPZ equation --- was a good model for surface propogation. More recently Hairer \cite{HAIRER13} applied deep results from probability theory to \textit{solve} the KPZ equation to much acclaim.

\subsection{Focus of this work: hierarchy and forced Burgers with time dependent coefficients}
Recently B\"{u}y\"{u}ka\c{s}ik and Pashaev \cite{BUYUKASIK13} and Schulze-Halberg \cite{HALBERG15} detailed a linearising transform of the Burgers equation with general forcing term and time dependent coefficients. Using results from these aforementioned studies, we focus on generating solutions to the Cauchy problem of the two following systems: 1) the Burgers hierarchy with general time dependent coefficients,
\begin{eqnarray}
\boxed{\frac{\partial}{\partial t}  U_N =(-1)^{\left\lfloor \frac{N+3}{2} \right\rfloor} \beta(t) \frac{\partial}{\partial z} \left( \frac{1}{\alpha(t)} \frac{\partial}{\partial z} - \frac{1}{2}U_N  \right)^N U_N  -\frac{\alpha'(t)}{\alpha(t)}  U_N},
\label{EQ1}
\end{eqnarray}
and 2) the forced Burgers equation with general time dependent coefficients,
\begin{eqnarray}
\boxed{ \frac{\partial}{\partial t}  U_{(F)} =\beta(t) \frac{\partial}{\partial z}\left( \frac{1}{\alpha(t)}\frac{\partial}{\partial z}U_{(F)} - \frac{1}{2}U^2_{(F)} +\frac{2 }{\alpha^2(t)} V(z) \right) -\frac{\alpha'(t)}{\alpha(t)}  U_{(F)}},
\label{EQ2}
\end{eqnarray}
where $N \in \mathbb{N}$ and both equations have the general initial condition $U_N(z,0)=U_{(F)}(z,0)= \psi(z)$. Specifically, $\alpha(t)$ and $\beta(t)$ are general coefficients differentiable in $t$, and $V(z)$ is an external spatial potential given explicitly in this work as,
\begin{eqnarray}
V(z) = \frac{(2 \gamma +\sigma_1)^2-1}{4 z^2}-\frac{(\sigma_1+1)\sigma_2}{2z} + \frac{\sigma^2_2}{4},
\label{potential}
\end{eqnarray}
for free variables $\{\gamma, \sigma_1, \sigma_2\}$. The potential in Eq.(\ref{potential}) is similar to those considered by Broadbridge \cite{BROADBRIDGE99} who applied the forced Burgers equation to model the rate plant roots extract water. Importantly, the corresponding linearised system of Eq.(\ref{EQ2}) is known as the \textit{sinked} Bessel process with constant drift, a generalisation of a system first solved by Linetsky \cite{LINETSKY04}. Additionally, the combination $\alpha'(t) U /\alpha(t)$ on the right hand side of both Eq.(\ref{EQ1}) and Eq.(\ref{EQ2}) act as positive/negative damping terms, depending on their sign. To the best of our knowledge, Eq.(\ref{EQ1}) is new in the literature, although it is \textit{similar} \cite{ POPOVYCH10} to Eq.(\ref{hierarch}) through transformations that we shall outline. Additionally, although Eq.(\ref{EQ2}) was first posed in \cite{HALBERG15} and solved for various potentials and initial conditions, in this work we derive the Green's function for the corresponding linearised system. 

\subsection{A remark on solution techniques}

As an analytic or numerical method to solve differential equations, we do not favour Green's functions over other equally valid methods. Indeed, we advocate the pursuit of \textit{multiple} solution techniques for the validation of modelling results; see \cite{Zuparic18} for a recent example. As a numerical tool in the applied sciences, one advantage of Green's functions lies in their ability to produce exact solution if one can apply integral identities, see \cite{GORSKA11} for an example involving Meijer-G function identities providing solution to L\'{e}vy noise probability densities. Furthermore, von Niessen \cite{vonNiessen91} reported Green's function solutions to multiple electron systems providing sufficient numerical accuracy and stability. Nevertheless Onida \textit{et al.} \cite{Onida02} gives an extremely nuanced account of the advantages and challenges posed by Green's functions as a means to model many electron-photon interactions, and argues for a solution approach tempered with multiple techniques.

We additionally remark that although we are concerned with the Cauchy problem of NLPDEs in an infinite/semi-infinite domain, various applications solve for equivalent boundary value problems in a finite/semi-infinite domain. Relevant examples include Broadbridge \textit{et al.} \cite{BROADBRIDGE97,BROADBRIDGE99} who considered the boundary value problem of various Burgers and related NLPDEs in finite domains to model unsaturated water in soil with plant roots extracting water. Furthermore, Calogero and De Lillo \cite{Calogero91} attempted to formulate a systematic approach to solving canonical boundary value problems for the conservative Burgers equation in the semi-infinte domain. For details on constructing Green's functions in finite domains refer to Linetsky \cite{Linetsky05}.

\subsection{Outline of paper}
In this work we solve the full Cauchy problem to the Burgers hierarchy, and forced Burgers equation, with time dependent coefficients. This is achieved by exploiting known solutions, and offering new expressions to the Green's functions of the corresponding linearly transformed systems, and using these to fully solve the associated NLPDEs. In the next section, we linearly transform Eq.(\ref{EQ1}) into a higher-order heat-type equation (HOHTE) using the appropriate Cole-Hopf identity. Using the results of G\'{o}rska \textit{et al.} \cite{GORSKA13}, we give the closed form solutions of the HOHTE Green's functions, for general $N$, as a convenient sum of generalized hypergeometric functions. In section \ref{SEC3} we linearly transform Eq.(\ref{EQ2}) into the corresponding Schr\"{o}dinger/Fokker-Plank equation with the potential $V(z)$ given in Eq.(\ref{potential}). Using Sturm-Liouville spectral classification \cite{LINETSKY04b, DUNFORD88, FULTON05} we construct the closed form solutions of the linear Green's function. In section \ref{SEC4}, given the relevant Green's functions of the corresponding linearly transformed systems, we plot various examples of solutions to the Cauchy problem of Eq.(\ref{EQ1}) and Eq.(\ref{EQ2}) which extend previous results. Finally, in section \ref{concandfut} we offer a discussion and detail future work.

\section{Linearising transformation I - Burgers hierarchy}
\label{SEC2}
In this section we offer analytic expressions to the Green's functions of the linearly transformed system of the Burgers hierarchy. To achieve this we exploit the work of G\'{o}rska \textit{et al.} \cite{GORSKA13}, effectively adding one more application to their significant result. We then apply these results in Section \ref{SEC4}, offering time dependent plots to the Cauchy problem of Eq.(\ref{EQ1}) of various $N$ values. We begin by considering the appropriate linear transformation technique.

\subsection{Higher-order heat-type equations}
Focusing on Eq.(\ref{EQ1}), we apply the generalised Cole-Hopf transform given in \cite{BUYUKASIK13, HALBERG15},
\begin{eqnarray}
U_N= - \frac{2}{\alpha (t)} \frac{\frac{\partial}{\partial z} \phi_N}{\phi_N}, 
\label{ColeHopf}
\end{eqnarray}
for proxy variable $\phi \equiv \phi (z,t)$, onto Eq.(\ref{EQ1}) to obtain the expression
\begin{eqnarray}
\begin{split}
\frac{\alpha^N(t)}{\beta(t)} \left( \frac{\frac{\partial}{\partial t} \phi_N}{\phi_N} \right) =  (-1)^{\left\lfloor \frac{N+3}{2} \right\rfloor} \left(  \frac{\partial}{\partial z} + \frac{\frac{\partial}{\partial z} \phi_N}{\phi_N}  \right)^N \frac{\frac{\partial}{\partial z} \phi_N}{\phi_N},\\
\Rightarrow \;\; \frac{\alpha^N(t)}{\beta(t)} \frac{\partial}{\partial t} \phi_N =  (-1)^{\left\lfloor \frac{N+3}{2} \right\rfloor}\frac{\partial^{N+1}}{\partial z^{N+1}} \phi_N ,
\label{intermediate}
\end{split}
\end{eqnarray}
where an integration constant has been set as zero \cite{KUDRYASHOV09b}. Proof that the second line of Eq.(\ref{intermediate}) follows from the first is given by Lemma 1 in Kudryashov and Sinelshchikov \cite{KUDRYASHOV1}. Moreover, constructing the auxiliary time variable $\tau_N$ as follows,
\begin{eqnarray}
\tau_N(t) = \int^t_{0} dq \frac{\beta(q)}{\alpha^N(q)},
\label{time}
\end{eqnarray}
transforms Eq.(\ref{intermediate}) into the canonical HOHTE 
\begin{eqnarray}
 \frac{\partial}{\partial \tau_N} \phi_N(z,\tau_N) =  (-1)^{\left\lfloor \frac{N+3}{2} \right\rfloor}\frac{\partial^{N+1}}{\partial z^{N+1}} \phi_N (z,\tau_N),
\label{HOHTE}
\end{eqnarray}
where the choice of coefficient of the spatial derivative ensures that the Green's functions of Eq.(\ref{HOHTE}) obey suitable arcsine laws \cite{HOCHBERG94}.

HOHTEs have themselves been the object of study for some time, an early example including \cite{HOCHBERG78} which explored the connection between HOHTEs of even order and generalised signed processes of unbounded measure. Other more recent examples involve HOHTEs as conditional probability densities associated with higher-order Kramers-Moyal equations generated from so-called polar noise \cite{DRUMMOND14}. Additionally, the Green's functions detailed in this work are similar to those featuring as governing equations to one-sided \cite{PENSON10} and two-sided \cite{GORSKA11, PIRYATINSKA05} L\'{e}vy stable distributions.

\subsection{Green's functions of higher-order heat-type equations}
In order to solve the Cauchy problem of Eq.(\ref{EQ1}), we are required to solve for the Green's function of the corresponding HOHTE, 
\begin{eqnarray}
 \frac{\partial}{\partial \tau} G_N(z,\tau_N|\xi) =  (-1)^{\left\lfloor \frac{N+3}{2} \right\rfloor}\frac{\partial^{N+1}}{\partial z^{N+1}} G_N (z,\tau_N|\xi), \;\; G_N (z,0|\xi) = \delta(z-\xi).
\label{GREEN1}
\end{eqnarray}
Taking the Fourier transform of Eq.(\ref{GREEN1}), solving the corresponding ordinary differential equation with respect to $\tau_N$, and then performing the inverse fourier transform we obtain,
\begin{eqnarray}
\begin{split}
G_N(z,\tau_N|\xi) &=& \frac{1}{2 \pi} \int^{\infty}_{-\infty}d \omega \exp\left[  i \omega (z-\xi)+ (-1)^{\left\lfloor \frac{N+3}{2} \right\rfloor } (i \omega)^{N+1} \tau_N \right],\\
&=& \left\{ \begin{array}{ll}
 \frac{1}{2 \pi} \int^{\infty}_{-\infty}d \omega \exp\left[  i \omega (z-\xi) - \omega^{N+1} \tau_N \right], & \textrm{ $N$ odd},\\
 \frac{1}{2 \pi} \int^{\infty}_{-\infty}d \omega \exp\left\{  i \left[ \omega (z-\xi)- \omega^{N+1} \tau_N \right] \right\}, & \textrm{ $N$ even}.
\end{array}
 \right. 
\end{split}
\label{inteq3}
\end{eqnarray}
Introducing the self-similar variables 
\begin{equation}
\chi= \tau^{1/(N+1)}_N\omega, \;\;  \zeta = \tau^{-1/(N+1)}_N(z-\xi) ,
\end{equation}
Eq.(\ref{inteq3}) results in the the Green's function
\begin{eqnarray}
\begin{split}
G_N(\zeta,\tau_N) = \left\{ \begin{array}{ll}
 \frac{1}{\pi \tau^{1/(N+1)}_N} \int^{\infty}_{0}d \chi \exp \left( -\chi^{N+1} \right) \cos (\chi \zeta ), & \textrm{ $N$ odd},\\
 \frac{1}{\pi \tau^{1/(N+1)}_N} \int^{\infty}_{0}d \chi \cos (\chi \zeta - \chi^{N+1} ),  & \textrm{ $N$ even}.
\end{array}
 \right. 
\end{split}
\label{prehyper}
\end{eqnarray}
The odd and even integrals are recognised as the \textit{symmetric L\'{e}vy stable signed functions} and \textit{generalised Airy functions} respectively. Both sets of integrals in Eq.(\ref{prehyper}) can be solved analytically via a Mellin and inverse Mellin transform to obtain compact Meijer-G function expressions---equations (23) and (32) of \cite{GORSKA13}. More convenient for computational purposes however is the Green's function as a finite sum of generalised hypergeometric functions \cite{Askey10},
\begin{equation}
G_N(\zeta,\tau_N) = \left\{ \begin{array}{cc} 
\sum^{\frac{N+1}{2}}_{j=1} \frac{ \tau^{-\frac{1}{(N+1)}}_N \kappa^{(o)}_{(N,j)}}{ \zeta^{2-2j}}  \,_0F_{N-1}\left(  \left. \begin{array}{c}
- \\
\tilde{\Delta} \left(N,2j \right)  \end{array}  \right| \left\{ \frac{-\zeta^2}{(N+1)^2} \right\}^\frac{N+1}{2} \right), & \textrm{$N$ odd},\\
\sum^{N}_{j=1} \frac{ \tau^{-\frac{1}{(N+1)}}_N \kappa^{(e)}_{(N,j)} }{ \zeta^{1-j} }   \,_0F_{N-1}\left(  \left. \begin{array}{c}
- \\
\tilde{\Delta} \left(N,1+j \right)  \end{array}  \right|  (-1)^{\frac{N}{2}}\left\{ \frac{\zeta}{N+1} \right\}^{N+1} \right), & \textrm{$N$ even},
\end{array}\right. 
\label{GREENSFUNC1}
\end{equation}
where $\tilde{\Delta} (l,k) = \{\frac{k}{l+1},\frac{k+1}{l+1}, \dots, \frac{k+l}{l+1}\}$ with entries $\{1,\frac{k+l}{l+1}\}$ deleted. The coefficients in Eq.(\ref{GREENSFUNC1}) are given by,
\begin{eqnarray}
\begin{split}
\kappa^{(o)}_{(N,j)} = \frac{\sqrt{\frac{N+1}{2 \pi}}}{(N+1)^{2j-1} \pi^{\frac{N+1}{2}}} \frac{\Gamma \left( \frac{N+2j}{N+1} \right) \prod^{N+1}_{\substack{k=1 \\ \ne 2 j}}\Gamma \left( \frac{k-2j}{N+1} \right) }{ \left\{ \prod^{\frac{N-1}{2}}_{k=0}\sin \left[ \pi \left( \frac{2 k - 2j +1}{N+1} \right) \right] \right\}^{-1}}, \\
\kappa^{(e)}_{(N,j)}= \frac{\sqrt{N+1}}{(N+1)^{j} \pi^{\frac{N+2}{2}}} \frac{\Gamma \left( \frac{N+1+j}{N+1} \right) \prod^{N+1}_{\substack{k=1 \\ \ne  j+1}}\Gamma \left( \frac{k-j-1}{N+1} \right) }{ \left\{ \prod^{\frac{N}{2}}_{k=0}\sin \left[ \pi \left( \frac{ 2k}{N+2}-\frac{j}{N+1} \right) \right] \right\}^{-1}},
\end{split}
\label{COEFF1}
\end{eqnarray}
where the superscripts $(o)$ and $(e)$ indicate \textit{odd} and \textit{even} respectively. Eqs.(\ref{GREENSFUNC1}) and (\ref{COEFF1}) provide the complete Green's function solution to the system in Eq.(\ref{GREEN1}), and are the main result of \cite{GORSKA13}.

\subsection{Example plots}
\begin{figure}
\begin{center}
\includegraphics[width=16cm]{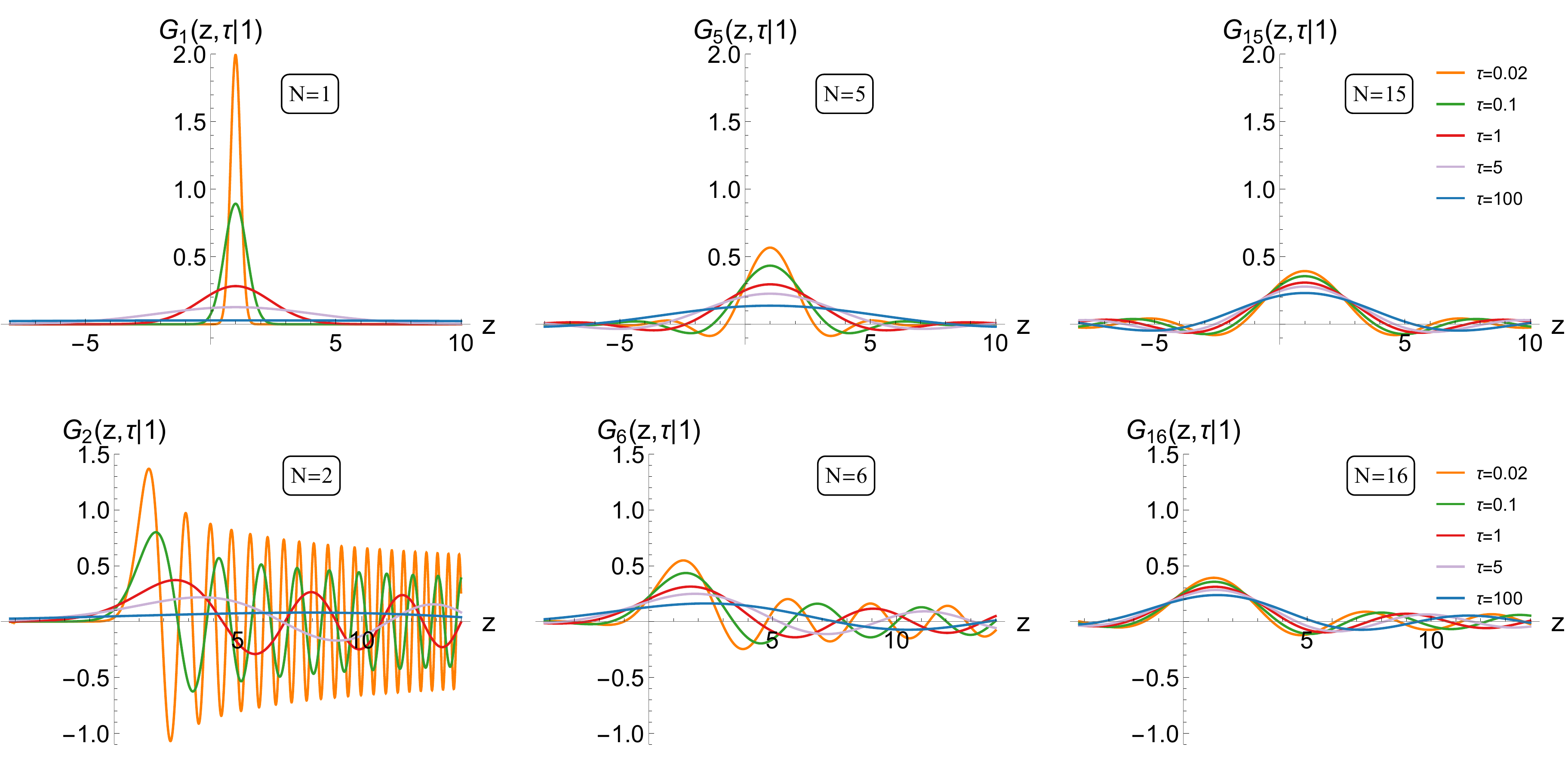}
\caption{Plots for Eq.(\ref{GREENSFUNC1}) for various instances of $N$, at different $\tau$ values and initial condition $\xi=1$. Note that odd/even values of $N$ are presented on the top/bottow row respectively.}
\label{fig:linearhierarchy}
\end{center}
\end{figure}

As mentioned previously, Kudryashov and Sinelshchikov \cite{KUDRYASHOV14} also considered the Cauchy problem of the Burger hierarchy, and were able to solve \textit{specific instances} of Eq.(\ref{prehyper}), obtaining the $N=\{1, 2, 3\}$ expressions of Eq.(\ref{GREENSFUNC1})---equations (31), (53) and (38) of \cite{KUDRYASHOV14} respectively. We plot various instances Eq.(\ref{GREENSFUNC1}) in Figure \ref{fig:linearhierarchy}. For the top row, which gives increasing values of odd-$N$, we note that the curves are symmetric around $z=1$, consistent with the initial condition $\xi=1$. $N=1$ is of course, the Gaussian solution of the heat-equation. In the bottom row of Figure \ref{fig:linearhierarchy} we give increasing values of even-$N$. For small-$N$, we note that the curves are not symmetric, highlighting the difference between the generalised Airy functions and the symmetric L\'{e}vy stable signed functions. However, we do see that the curves for $N=15$ and $16$ are very similar, reflecting the observation in \cite{GORSKA13} that for large $N$, the difference between even and odd instances becomes negligible. Importantly, we have performed a simple validation of Eq.(\ref{GREENSFUNC1}) and (\ref{COEFF1}) by reproducing figures $1, 2$ and $3$ in \cite{GORSKA13} with the expression $G_N(z,1|0)$.

\subsection{General solution to the Cauchy problem}

Applying the Green's functions, the solution to Eq.(\ref{EQ1}) with the general initial condition $U(z,0) = \psi(z)$, and general order $N$ is given as,
\begin{eqnarray}
U_N(z,t) = - \frac{2}{\alpha(t)}\frac{\frac{\partial}{\partial z} \int^{\infty}_{-\infty} d\xi G_{N}(z,\tau_N(t)|\xi) \exp \left[ -\frac{\alpha(0)}{2} \int^{\xi} d \chi \psi(\chi) \right]}{ \int^{\infty}_{-\infty} d\xi G_{N}(z,\tau_N(t)|\xi) \exp \left[ -\frac{\alpha(0)}{2}\int^{\xi} d \chi \psi(\chi) \right]} ,
\label{bigsol1}
\end{eqnarray}
for $G_{N}(z,\tau_N|\xi)$ given by Eq.(\ref{GREENSFUNC1}). We shall consider numerical examples of Eq.(\ref{bigsol1}) in section \ref{SEC4} to solve the Cauchy problem presented by Eq.(\ref{EQ1}).

\section{Linearising transformation II - forced Burgers equation}
\label{SEC3}
\subsection{Schr\"{o}dinger and Fokker-Planck equations - sinked Bessel process with constant drift}

Focusing now on Eq.(\ref{EQ2}), we apply the same transformations given in Eqs.(\ref{ColeHopf}) and (\ref{time}),
\begin{equation}
U_{(F)}= - \frac{2}{\alpha (t)} \frac{\frac{\partial}{\partial z} \phi_{(F)}}{\phi_{(F)}}, \;\; \tau(t) = \int^t_{0} dq \frac{\beta(q)}{\alpha(q)}, 
\end{equation}
to obtain,
\begin{eqnarray}
\frac{\partial}{\partial \tau} \phi_{(F)}(z,\tau) = \left( \frac{\partial^2}{\partial z^2} - V(z) \right) \phi_{(F)}(z,\tau),
\end{eqnarray}
which is the non-complex Schr\"{o}dinger equation. As mentioned recently by Yadav \cite{YADAV1}, despite its linearisability, exact series solutions to the forced Burgers equation are rare in the literature. Hence we devote the current section to deriving the closed form solution to the Green's function of the aforementioned Schr\"{o}dinger equation, namely
\begin{eqnarray}
 \frac{\partial}{\partial \tau} G_{(F)}(z,\tau|\xi) =  \left( \frac{\partial^2}{\partial z^2} - V(z) \right) G_{(F)} (z,\tau|\xi), \;\; G_{(F)}(z,0|\xi) = \delta(z-\xi).
\label{GREEN2}
\end{eqnarray}
This is achieved through the Schr\"{o}dinger equation's connection to the corresponding density evolution equation,
\begin{eqnarray}
\frac{\partial}{\partial \tau} X(x,\tau|y) = \left(\frac{\partial^2}{\partial x^2}s(x) - \frac{\partial}{\partial x}q(x) -r(x) \right) X(x,\tau|y), 
\label{FP} 
\end{eqnarray}
with initial condition $X(x,0|y) = \delta(x-y)$. Coefficients $r(x), q(x)$ and $s(x)$ are referred to as the \textit{sink}, \textit{drift} and \textit{diffusion} respectively. With $r(x)=0$, Eq.(\ref{FP}) is conservative and is generally referred to as the Fokker-Planck equation. In this work we consider coefficients with the following forms
\begin{eqnarray}
s(x) = 1, \;\; q(x) = \frac{\sigma_1+1}{x}-\sigma_2, \;\; r(x) = \frac{\gamma(\gamma+\sigma_1)}{x^2}.
 \label{densitycoeff}
\end{eqnarray}
This choice is referred to as the sinked Bessel process with constant drift, and has broad applicability: for zero sink these include queueing theory in the heavy traffic limit \cite{Coffman98}, the biological process of \textit{DNA breathing} \cite{Fogedby07}, and the distribution of Arctic sea ice thickness \cite{Toppaladoddi15}. The appearance of the sink coefficient $r(x)$ additionally has application in finance, where it is referred to as the \textit{instantaneous discount rate} \cite{LINETSKY04, LINETSKY04b, Linetsky04c}.

Importantly, Eq.(\ref{GREEN2}) and Eq.(\ref{FP}) are linked through the \textit{Liouville transformation}:
\begin{eqnarray}
\xi(x) = z(x) = \int^x \frac{d \zeta}{\sqrt{s(\zeta)}}, \;\; X(x,\tau|y) = \frac{G_{(F)}(z(x),\tau| \xi(y))}{ \sqrt{\frac{W(y)}{ W(x)} \sqrt{s(x)s(y)} }}.
\label{liouville1}
\end{eqnarray}
As $s(x) = 1$, we simply obtain $z(x) =x$. The expression $W(x)$ in Eq.(\ref{liouville1}) is commonly referred to as the \textit{weight function}, and is the solution to the \textit{Pearson} equation:
\begin{equation}
\begin{array}{rcll}
W(x) &=& \frac{\kappa}{s(x)}\exp \left[ \int^xd\zeta \frac{q(\zeta)}{s(\zeta)} \right], &\kappa \in \mathbb{R},\\
& = &x^{\sigma_1+1}e^{-\sigma_2 x}, & \{x, \sigma_2 \} \in \mathbb{R}_+.
\end{array}
\label{weightF}
\end{equation}
As detailed in \cite{LINETSKY04b,FULTON05}, the solution to Eq.(\ref{FP}) is given by the general expression
\begin{eqnarray}
X(x,\tau|y)= W(x)\SumInt_{\lambda} e^{-\lambda \tau}\rho_{\lambda} \vartheta_{\lambda} (x) \vartheta_{\lambda} (y), \;\; \rho_{\lambda} \in \mathbb{C},
\label{genX}
\end{eqnarray}
where a sum is applied for discrete eigenspectra, and an integral for corresponding continuous eigenspectra. The Liouville transformation plays an integral role in determining the spectral properties of Eq.(\ref{genX}). 

More generally, the solutions that we detail here fit into a class of Schr\"{o}dinger potentials/density coefficients which are \textit{exactly solvable} \cite{Turbiner88}. That is, each eigenfunction can be expressed as a closed form hypergeometric function, as opposed to something more generally transcendental \cite{Derezinski11}. For the derivation of Green's functions associated with other exactly solvable sinked density equations, refer to \cite{ZUPARIC15}.

The remainder of this section is devoted to the derivation of the analytical form of Eq.(\ref{genX}) using a combination of the spectral classification found in Linetsky \cite{LINETSKY04b} and the application of an inverse integral transformation known as a MacRobert's proof; the main benefit of this approach is that it solely relies on the asymptotic properties of the corresponding eigenfunctions, and knowledge of Dirichlet integrals. Analytical expressions of the conservative (\textit{i.e.} $r(x)=0$) Bessel process with constant drift were first found using complex variable approaches to spectral expansions \cite{LINETSKY04}, and more recently in \cite{Guarnieri17} who employ a technique similar to the method of stationary phase; we extend these results by deriving the corresponding \textit{sinked} ($r(x) \ne 0$) expressions.

\subsection{Self-adjoint Sturm-Liouville operators}
Decomposing $X(x,\tau|y)$ into the weight function $W(x)$, multiplied by an auxiliary function $Y(x,\tau|y)$ --- i.e. $X(x,\tau|y)= W(x) Y(x,\tau|y)$ --- Eq.(\ref{FP}) for the auxiliary function becomes,
\begin{eqnarray*}
\frac{\partial}{\partial \tau} Y(x,\tau|y) = \underbrace{\left(s(x)\frac{\partial^2}{\partial x^2} + q(x)\frac{\partial}{\partial x} -r(x) \right)}_{\equiv {\cal H}} Y(x,\tau|y),
\end{eqnarray*}
where ${\cal H}$ is referred to as the Sturm-Liouville operator. In order to ensure finiteness of solutions we require that all eigenvalues of ${\cal H}$ are negative:
\begin{eqnarray}
{\cal H}  \vartheta_{\lambda} (x) = - \lambda  \vartheta_{\lambda} (x), \;\; \lambda \ge 0.
\label{S-L}
\end{eqnarray}
For all self-adjoint, non-positive ${\cal H}$ the set of solutions to Eq.(\ref{S-L}) forms a weighted square integrable Hilbert space \cite{McKean56}, $L^2 (\{x:W(x)>0\})$, with respect to the inner product,
\begin{eqnarray*}
\int_{W(x)>0}dx W(x) \vartheta_{\lambda} (x)\vartheta_{\mu} (x) < \infty.
\end{eqnarray*}
Additionally, if the spectrum to Eq.(\ref{S-L}) is mixed (discrete and continuous), the Hilbert space is separable into the following orthogonal subspaces
\begin{eqnarray*}
L^2 _{pp}(\{x:W(x)>0\}) \oplus L^2_{ac} (\{x:W(x)>0\}),
\end{eqnarray*}
for $L^2_{pp}$ denoting the \textit{pure point} Hilbert subspace and $L^2_{ac}$ denoting the \textit{absolutely continuous} Hilbert subspace. Finally, if we label the eigenfunctions corresponding to each Hilbert subspace as
\begin{eqnarray*}
\vartheta_{n} (x) \in L^2 _{pp}(\{x:W(x)>0\}),\;\; n \in \{ 0,1,\dots\}, \\
\vartheta(\mu, x) \in L^2 _{ac}(\{x:W(x)>0\}) , \;\; \mu > 0,
\end{eqnarray*} 
then by the orthogonality condition we have
\begin{equation}
\int_{W(x)>0}dx W(x) \vartheta_{n} (x)\vartheta_{m} (x) = \frac{\delta_{nm}}{\rho_n} \;\;\textrm{and} \;\; \int_{W(x)>0}dx W(x) \vartheta_{n} (x)\vartheta (\mu,x) = 0 .
\label{orthog1}
\end{equation}

\subsection{Schr\"{o}dinger potential and the eigenspectrum}
Under the Liouville transformation of Eq.(\ref{liouville1}), the density coefficients and the Schr\"{o}dinger potential of Eq.(\ref{potential}) are explicitly related by
\begin{eqnarray}
V(z) = \frac{3 \left[ s'(x) \right]^2}{16 s(x)} -\frac{s''(x)}{4} + \frac{q^2(x)}{4 s(x)} + 
\frac{q'(x)}{2} - \frac{q(x)s'(x) }{2 s(x)}+r(x),
\end{eqnarray}
for $x \equiv x(z)$ in the above expression. Following \cite{LINETSKY04b}, by examining the properties of Eq.(\ref{potential}) and determining if the endpoints of the domain of $x$ are \textit{oscillatory/non-oscillatory}, we can determine all the spectral properties ${\cal H}$. Our domain of $x$, as indicated in Eq.(\ref{weightF}) is given by $\mathbb{R}_+$. For our particular example of ${\cal H}$, we obtain that it is classified as non-oscillatory at the domain endpoint $x=0$.

For the situation $\lim_{x \rightarrow \infty} z(x) = \infty$ and $\lim_{x \rightarrow \infty} V(x) = \Lambda < \infty$, ${\cal H}$ is classified as oscillatory/non-oscillatory at the domain endpoint $x=\infty$, with \textit{cutoff} at $\Lambda$. Additionally, for $\lim_{x \rightarrow \infty} z^2(x) (V(x) -\Lambda) < -1/4$, the spectrum for ${\cal H}$ is non-oscillatory for $\lambda \in [0,\Lambda)$, and oscillatory for $\lambda \ge \Lambda$.

Moreover, given the Schr\"{o}dinger potential in Eq.(\ref{potential}) we obtain
\begin{eqnarray}
\Lambda = \frac{\sigma^2_2}{4} < \infty, \;\; \textrm{and, } \;\;\lim_{x \rightarrow \infty} z^2(x) (V(x) -\Lambda) = -\infty  < -\frac{1}{4}.
\end{eqnarray}
Thus for this particular Sturm-Liouville operator, the endpoint at zero is classed as non-oscillatory, and the endpoint at $\infty$ is oscillatory/non-oscillatory. Specifically, the spectrum is non-oscillatory for $\lambda \in [0,\sigma^2_2/4)$, and oscillatory for $\lambda \ge \sigma^2_2/4$.
\subsection{Complete density solution and example plots}
 Following \cite{LINETSKY04,LINETSKY04b}, the spectral classification implies that the solution to Eq.(\ref{FP}) has the specific form,
\begin{equation}
X(x,\tau|y) =W(x) 
\left[
\sum^{\infty}_{n=0}e^{-\lambda_n \tau} \rho_n \vartheta_n (x) \vartheta_n (y)+ \int^{\infty}_{0}d \mu e^{-\left(\frac{\sigma^2_2}{4}+\mu^2\right)\tau} \rho(\mu) \vartheta (\mu,x) \vartheta (\mu,y) 
\right],\label{solX}
\end{equation}
where the discrete eigenvalues $\lambda_n$ approaching $\sigma^2_2/4$ from the left become dense. For convenience we provide the full list of quantities for Eq.(\ref{solX}) in Table \ref{tab:tab1}. 
\begin{table}[ht]
\caption{Explicit expressions of relevant quantities in Eq.(\ref{solX}) --- variables $a$ and $b$ are defined in Eq.(\ref{modifiedS-L}).} 
\centering 
\begin{tabular}{c c c } 
\hline 
Quantity & Expression & Restrictions \\ [0.5ex] 
\hline 
$W(x)$ & $x^{\sigma_1+1}e^{-\sigma_2 x}$ & $\{x,\sigma_2\} > 0$  \\
$\lambda_n$ & $\frac{(n+\gamma)(n+2b-\gamma)\sigma^2_2}{4(n+b)^2} $ & $\gamma (2 b - \gamma) \ge 0$  \\
$\rho_n$ & $ \frac{n! \left( \frac{2a}{n +b} \right)^{2b +1}}{2(n +b)\Gamma(n +2b)}$ & $b \ne - \mathbb{N}_+ \cup 0$   \\
$ \vartheta_{n} (x)$ & $x^{\gamma} e^{  \frac{ (n+\gamma)\sigma_2}{2(n+b)}  x } L^{(2b-1)}_n \left( \frac{2a}{ n +b }x \right) $ & $b >0, \;\;\gamma \ge 0$  \\
$\rho(\mu)$ & $  \frac{(2 \mu)^{2b} e^{\frac{\pi a}{\mu}} \left| \Gamma\left( b+ i \frac{a}{\mu} \right) \right|^2}{2 \pi} $ & $\mu > 0$   \\
$ \vartheta_{a,b} (\mu,x)$ & $ \frac{ x^{\gamma} e^{ \left(\frac{\sigma_2}{2} - i \mu\right) x }}{\Gamma(2b)} 
\,_1F_1\left( \left. \begin{array}{c}
b+i \frac{a}{ \mu} \\
2b \end{array}  \right| i2\mu x \right)$ & $\mu > 0$  \\ [1ex] 
\hline 
\end{tabular}
\label{tab:tab1} 
\end{table}

The restrictions in Table \ref{tab:tab1} are to ensure that the eigenfunctions form a weighted square integrable Hilbert space. The discrete and continuous spectrum normalisations are determined by manipulation of the initial condition expression,
\begin{eqnarray}
\delta(x-y) =W(x) 
\left[
\sum^{\infty}_{n=0} \rho_n \vartheta_n (x) \vartheta_n (y)+ \int^{\infty}_{0}d \mu  \rho(\mu) \vartheta (\mu,x) \vartheta (\mu,y) 
\right],
\label{InitialC}
\end{eqnarray}
and application of the orthogonality conditions given in Eq.(\ref{orthog1}). The derivation of eigenvalues, eigenfunctions and normalisations associated with Eq.(\ref{solX}) are presented in the following subsections.

\begin{figure}
\begin{center}
\includegraphics[width=11.0cm]{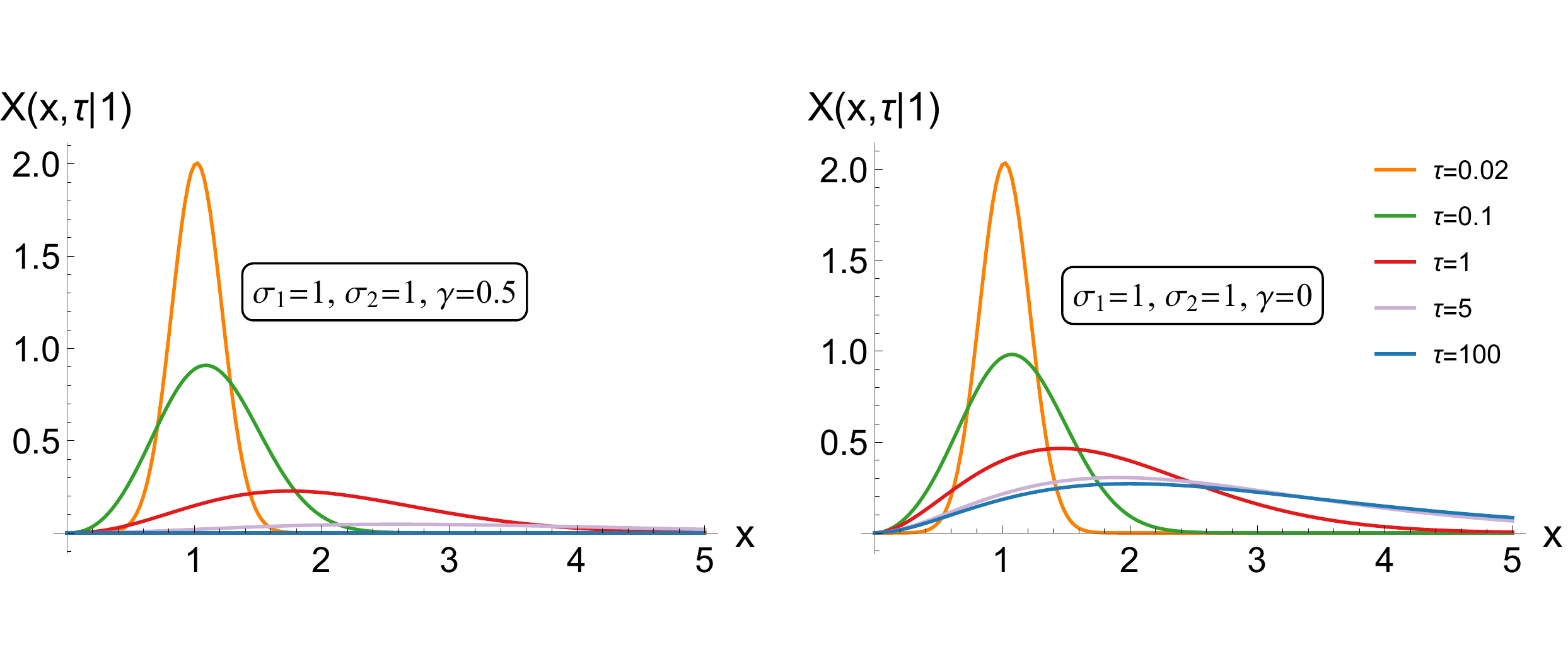}
\caption{Plots for Eq.(\ref{solX}) for parameter values $\sigma_1 = \sigma_2 = 1$, initial condition $y=1$, at different $\tau$ values. The sink parameter $\gamma$ for the left and right panels is given as $0.5$ and $0$ respectively.}
\label{fig:linearforced}
\end{center}
\end{figure}

We give example plots of Eq.(\ref{solX}) in Figure \ref{fig:linearforced} for various time values with parameters $\sigma_1 = \sigma_2 = 1$ and initial condition $y=1$. In the left panel we have set $\gamma = 0.5$, meaning that the density is indeed sinked. This results in the density profiles for $\tau \ge 5$ appearing negligible compared to earlier times, given the absence of conservation. Contrastingly, in the right panel we give the corresponding density profiles for the conservative case ($\gamma = 0$).  We observe that for small $\tau$-values the profiles in the right panel look similar to those on the left, whereas for $\tau \ge 1$, the profiles on the left panel are suppressed compared to the conservative case on the right.

\subsection{Confluent hypergeometric differential equation}
If we decompose the eigenfunction $ \vartheta_{\lambda} (x)$ in Eq.(\ref{S-L}) as
\begin{eqnarray*}
\vartheta_{\lambda} (x) = x^{\gamma} \exp \left( \frac{ \sigma_2-\sqrt{\sigma^2_2 - 4\lambda}}{2}  x \right) \varphi_{\lambda}(x), 
\end{eqnarray*}
and transform the variable $x$ as $u= x \sqrt{\sigma^2_2-4 \lambda}$, the corresponding Sturm-Liouville expression for $\varphi_{\lambda}(u)$ becomes,
\begin{eqnarray}
\begin{split}
\left[u \frac{d^2}{du^2} +\left(2b - u \right) \frac{d}{du} + \left( \frac{2a}{\sqrt{\sigma^2_2 - 4\lambda}} - b \right) \right]\varphi_{\lambda}(u) =0, \label{modifiedS-L} \\
\textrm{  where  }\;\;a = \frac{(\sigma_1+1)\sigma_2}{4}, \;\; b = \frac{\sigma_1+2\gamma+1}{2}.
\end{split}
\end{eqnarray}
Following \cite{STEGUN} we recognise Eq.(\ref{modifiedS-L}) as the confluent hypergeometric differential equation; depending on the particular choice of eigenvalue, the eigenfunction can be either the $n$th order \textit{Laguerre polynomial} or the more general \textit{confluent hypergeometric function}. We consider first the polynomials for the discrete part of the spectrum.

\subsection{Discrete spectrum - Laguerre polynomials}

Inserting the following expression for the eigenvalue
\begin{eqnarray}
\lambda_n = \frac{(n+\gamma)(n+2b-\gamma)\sigma^2_2}{(2n+2b)^2}\;\; \textrm{for} \;\; n \in \{0,1,\dots\},
\end{eqnarray}
transforms Eq.(\ref{modifiedS-L}) into the $n$th order Laguerre polynomial differential equation \cite{Koekoek}, with eigenfunction
\begin{eqnarray}
\begin{split}
\varphi_{n}(u) = L^{(2b-1)}_n(u) = \frac{(2b)_n}{n!}  \,_1F_1\left( \left. \begin{array}{l}
-n\\
2b \end{array}  \right| u \right),\\
\Rightarrow \vartheta_{n} (x) = x^{\gamma} e^{  \frac{ (n+\gamma)\sigma_2}{2n+2b}  x } L^{(2b-1)}_n \left( \frac{2a}{ n +b }x \right) .
\end{split}
\end{eqnarray}
To derive the corresponding discrete spectrum normlisation coefficients, we multiply both sides of Eq.(\ref{InitialC}) by $\vartheta_m(x)$, integrate over all $x$ and apply the orthogonality conditions in Eq.(\ref{orthog1}) to obtain
\begin{eqnarray}
\vartheta_m(y) = \rho_n  \vartheta_n(y)\int^{\infty}_0 dxW(x) \vartheta_m (x) \vartheta_n (x).
\end{eqnarray}
Additionally, applying an inductive proof it is possible to show that 
\begin{eqnarray}
\int^{\infty}_0 dxW(x) \vartheta_m (x) \vartheta_n (x) = \delta_{mn} \frac{(2m +2b)\Gamma(m +2b)}{m! \left( \frac{2a}{m +b} \right)^{2b +1}},
\end{eqnarray}
thus leading to the form of the normalisation $\rho_n$ given in Table \ref{tab:tab1}.

\subsection{Continuous spectrum - Confluent hypergeometric functions}

Inserting the eigenvalue expression, $\lambda = \sigma^2_2/4+\mu^2 $ into Eq.(\ref{modifiedS-L}), the eigenfunction becomes the more general confluent hypergeometric function $\varphi_{a,b}( \mu,u)$ given by,
\begin{eqnarray}
\begin{split}
\varphi_{a,b}( \mu,u)= \frac{1}{\Gamma(2b)} \,_1F_1\left( \left. \begin{array}{c}
b+i \frac{a}{ \mu}  \\
2b \end{array}  \right| u \right),\\ 
\Rightarrow \vartheta_{a,b} (\mu, x) = \frac{ x^{\gamma} e^{ \left(\frac{\sigma_2}{2} - i \mu\right) x }}{\Gamma(2b)} 
\,_1F_1\left( \left. \begin{array}{c}
b+i \frac{a}{ \mu} \\
2b \end{array}  \right| i2\mu x \right).
\end{split}
 \label{efunc2}
\end{eqnarray}
In an equivalent process to the discrete case, to derive the continuous spectrum normlisation coefficient, we multiply both sides of Eq.(\ref{InitialC}) by $\vartheta_{a,b}(\nu,x)$, integrate over all $x$ and apply the orthogonality conditions in Eq.(\ref{orthog1}) to obtain
\begin{eqnarray}
\vartheta_{a,b}(\nu,y) =\underbrace{ \int^{\infty}_0 dx W(x) \vartheta_{a,b}(\nu,x)  \int^{\infty}_0 d\mu \rho(\mu) \vartheta_{a,b}(\mu,x) \vartheta_{a,b}(\mu,y)}_{ \equiv {\cal I}(\nu,y)}.
\label{MacRob}
\end{eqnarray}
The process of isolating $\rho(\mu)$ in the integral expression in Eq.(\ref{MacRob}) is more commonly referred to as a \textit{MacRobert's proof} \cite{MACROBERT, Davies02}, in doing so we obtain the following expression for the continuous spectrum normalisation
\begin{equation}
\rho(\mu) = \frac{(2 \mu)^{2b} e^{\frac{\pi a}{\mu}} \left| \Gamma\left( b+ i \frac{a}{\mu} \right) \right|^2}{2 \pi},
\label{contnorm}
\end{equation}
where we have relegated the details to \ref{Appendix1}.

\subsection{General solution to the Cauchy problem}

Similar to the expression given in Eq.(\ref{bigsol1}) for the solution to the Burgers hierarchy, the solution to Eq.(\ref{EQ2}) for the same initial condition $(U_{(F)}(z,0)= \psi(z))$ is given as,
\begin{eqnarray}
U_{(F)}(z,t) = - \frac{2}{\alpha(t)}\frac{\frac{\partial}{\partial z} \int_{W(y(\xi)) > 0} d\xi G_{(F)} (z,\tau(t)|\xi) \exp \left[ -\frac{\alpha(0)}{2} \int^{\xi} d \chi \psi(\chi) \right]}{ \int_{W(y(\xi)) > 0} d\xi G_{(F)}(z,\tau(t)|\xi) \exp \left[ -\frac{\alpha(0)}{2} \int^{\xi} d \chi \psi(\chi) \right]},
\label{bigsol2}
\end{eqnarray} 
where from Eq.(\ref{liouville1}) we have,
\begin{eqnarray}
G_{(F)}(z,\tau| \xi) =  \sqrt{\frac{W(y(\xi))}{ W(x(z))} \sqrt{s(x(z))s(y(\xi))} }  X(x(z),\tau|y(\xi)),
\end{eqnarray}
and $X(x,\tau|y)$ is given by Eq.(\ref{solX}) and Table \ref{tab:tab1}. To the best of our knowledge the time dependent density offered in this section is an entirely new result, which we shall use in the Section \ref{SEC4} to construct solutions to Eq.(\ref{EQ2}).
\section{Numerical examples}
\label{SEC4}
\subsection{Burgers hierarchy}
\begin{figure}
\begin{center}
\includegraphics[width=16cm]{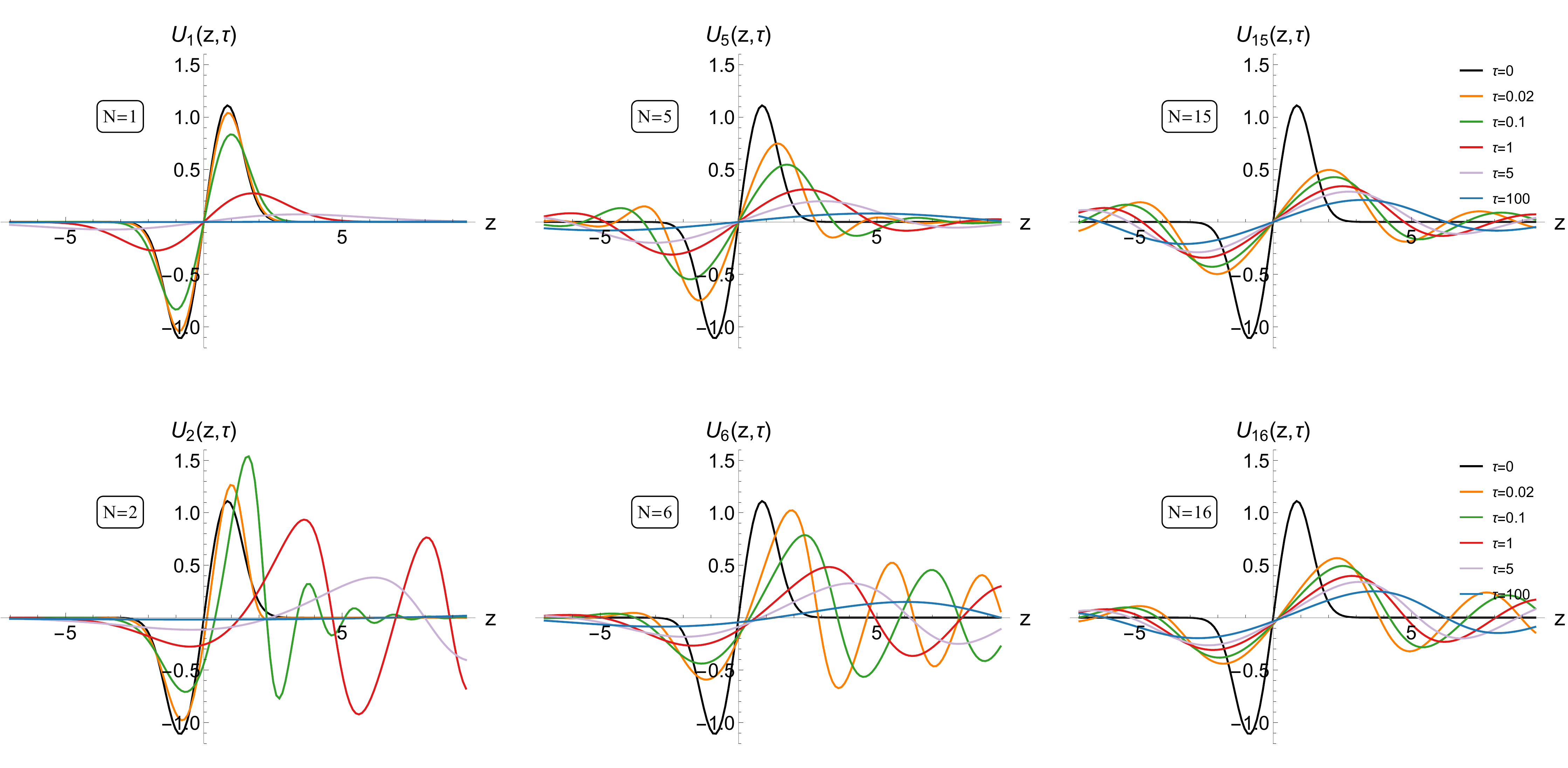}
\caption{Plots for Eq.(\ref{dampCauchyhie}) with initial condition $\psi(z)$ given by Eq.(\ref{inihierach}). Each panel, which gives a different instance of $N$, at different $\tau$ values, is presented to match the $N$ and $\tau$ values given in figure \ref{fig:linearhierarchy}. Note that odd/even values of $N$ are presented on the top/bottow row respectively.}
\label{fig:hierarchy-burgers}
\end{center}
\end{figure}

Setting $\alpha(t) = \beta(t) = 1$, (and hence $\tau_N(t)=t$) we choose the initial condition
\begin{eqnarray}
\psi(z) = \frac{4 z e^{-z^2}}{1+e^{-z^2}}\;\; \Rightarrow\;\;  \exp \left[ -\frac{\alpha(0)}{2} \int^{z} d \chi \psi(\chi) \right] =  1+e^{-z^2},
\label{inihierach}
\end{eqnarray}
which is the same as the initial condition chosen in \cite{KUDRYASHOV14}, except for a multiplicative factor. Substituting Eq.(\ref{inihierach}) into the denominator of Eq.(\ref{bigsol1}) we conveniently obtain
\begin{eqnarray}
\begin{split}
\int^{\infty}_{-\infty} d\xi G_{N}(z,t|\xi) \exp \left[ -\frac{1}{2}\int^{\xi} d \chi \psi(\chi) \right] = 1 + \int^{\infty}_{-\infty} d\xi G_{N}(z,t|\xi) e^{-\xi^2},\\
\Rightarrow \;\; U_N (z,t) = -2\frac{\frac{\partial}{\partial z} \int^{\infty}_{-\infty} d\xi G_{N}(z,t|\xi) e^{-\xi^2}}{1 + \int^{\infty}_{-\infty} d\xi G_{N}(z,t|\xi) e^{-\xi^2}} ,
\end{split}
\label{dampCauchyhie}
\end{eqnarray}
allowing for an exponential damping term which is especially important for the case $N=2$ where the Green's function can be expressed as a single Airy function, which itself has no damping properties for a negative argument. We give example plots of Eq.(\ref{dampCauchyhie}) in Figure \ref{fig:hierarchy-burgers}, where each of the six panels are presented to match the $N$ and $\tau$ values given in Figure \ref{fig:linearhierarchy}. 

In Figure \ref{fig:hierarchy-burgers} we notice that the top panels (for odd-$N$) show antisymmetric behaviour in the time evolution dynamics about the origin, reflecting the antisymmetry of the initial condition. We see similar behaviour being displayed by the $N=3$ case in Figure 3 of \cite{KUDRYASHOV14}. Comparing all three top panels of Figure \ref{fig:hierarchy-burgers}, we see that for low-$N$ the profiles experience significant damping as $|z|$ increases. Conversely, as $N$ increases, this damping lessens with increasing $|z|$, with the profiles displaying more oscillations. Focusing on the bottom panels (for even-$N$) of Figure \ref{fig:hierarchy-burgers}, the low-$N$ profiles are no longer antisymmetric---instead reflecting the complete asymmetry in the corresponding Green's functions. As $N$ increases, the asymmetry displayed at the origin on the bottom panels of Figure \ref{fig:hierarchy-burgers} lessens. Indeed, we see that the right-most plots of Figure \ref{fig:hierarchy-burgers} are qualitatively very similar, echoing the previous observation that for large-$N$, the difference between even and odd Green's functions becomes negligible \cite{GORSKA13}. Furthermore, we notice that as $N$ increases, it appears that it takes significantly more time-units for the profiles to reach steady-state behaviour (flat-line at $U_N=0$).

\subsection{A remark on numerical computation}
\label{numcomp}
The numerical integration function \texttt{NIntegrate} of \textit{Mathematica}\textsuperscript{\textregistered} 11.1.1.0 was utilised to directly evaluate the integrals of Eq.(\ref{dampCauchyhie}) for the Burgers hierarchy. The various integrals being evaluated were polynomials in hypergeometric functions, which having been obtained from the Green's function necessarily asymptotically decayed to zero. Nonetheless, the modulus of each individual term in these polynomials grew very fast as their argument tended to infinity, and the failure of these very large factors to asymptotically cancel due to finite machine precision errors lead---for very large bounds of integration---to very large numerical errors in each integrand. As the argument of each integrand grew the asymptotic decay to zero was eventually observed to suddenly transition into a regime of extremely rapid oscillation with at least exponentially increasing amplitude. Thus, these regions of extreme numerical error were able to be identified by inspection. Furthermore, virtually every attempt to numerically integrate over a poor choice of integration bounds where such numerical failures occurred caused our software package to emit a ``Numerical integration converging too slowly'' or ``\texttt{NIntegrate} failed to converge to prescribed accuracy...'' warning.

Careful attention was paid to specifying finite bounds of integration which avoided these extremal regions where catastrophic loss of numerical precision occurred, in order that a negligible error was introduced into the final integral calculation. As the most appropriate choice of bounds varied as a function of both $z$ and $t$, an automated preprocessing step of identifying these bounds---prior to the actual computation of the integrals---was adopted for each $z$ value. In particular, the oscillatory behavior of the integrands as $|\xi|$ becomes large was exploited, with a numerical routine counting an adjustable number of sign changes of the integrand before cutting the integral off in a given direction. Permitting a larger number of sign changes in preprocessing increased the precision of the integral, at the expense of encountering errors caused by loss of numerical precision of the generalised hypergeometric expressions for large $|\xi|$ values. For our purposes, it was found that a cutoff after \textit{five} sign changes was sufficient.

\subsection{Forced Burgers equation}
We illustrate our solutions to the Cauchy problem of the forced Burgers equation by considering the following more general time dependent forms for $\alpha$ and $\beta$,
\begin{eqnarray}
\begin{split}
\alpha(t) = \frac{1}{2} \left(1+ e^{-t} +\frac{1}{2}\sin \pi t  \right), \;\; \beta(t) = 1- \frac{1}{2}e^{-t},\\
\Rightarrow \tau (t) = \int^t_0 dq \frac{2-e^{-q}}{1+e^{-q}+ \frac{1}{2}\sin \pi q} ,
\end{split}
\end{eqnarray}
hence $\alpha(0) =1$. Choosing the constant initial condition $\psi(z) =2$, results in the intergrands in Eq.(\ref{bigsol2}) being exponentially suppressed via
\begin{eqnarray}
\begin{split}
\exp \left[ -\frac{\alpha(0)}{2} \int^{z} d \chi \psi(\chi) \right] = e^{-z},\\
\Rightarrow \;\; U_{(F)} (z,t) = -\frac{2}{\alpha (t)}\frac{\frac{\partial}{\partial z} \int^{\infty}_{0} d\xi G_{(F)}(z,t|\xi) e^{-\xi}}{\int^{\infty}_{0} d\xi G_{(F)}(z,t|\xi) e^{-\xi}} .
\end{split}
\label{force-burg}
\end{eqnarray} 

\begin{figure}
\begin{center}
\includegraphics[width=12cm]{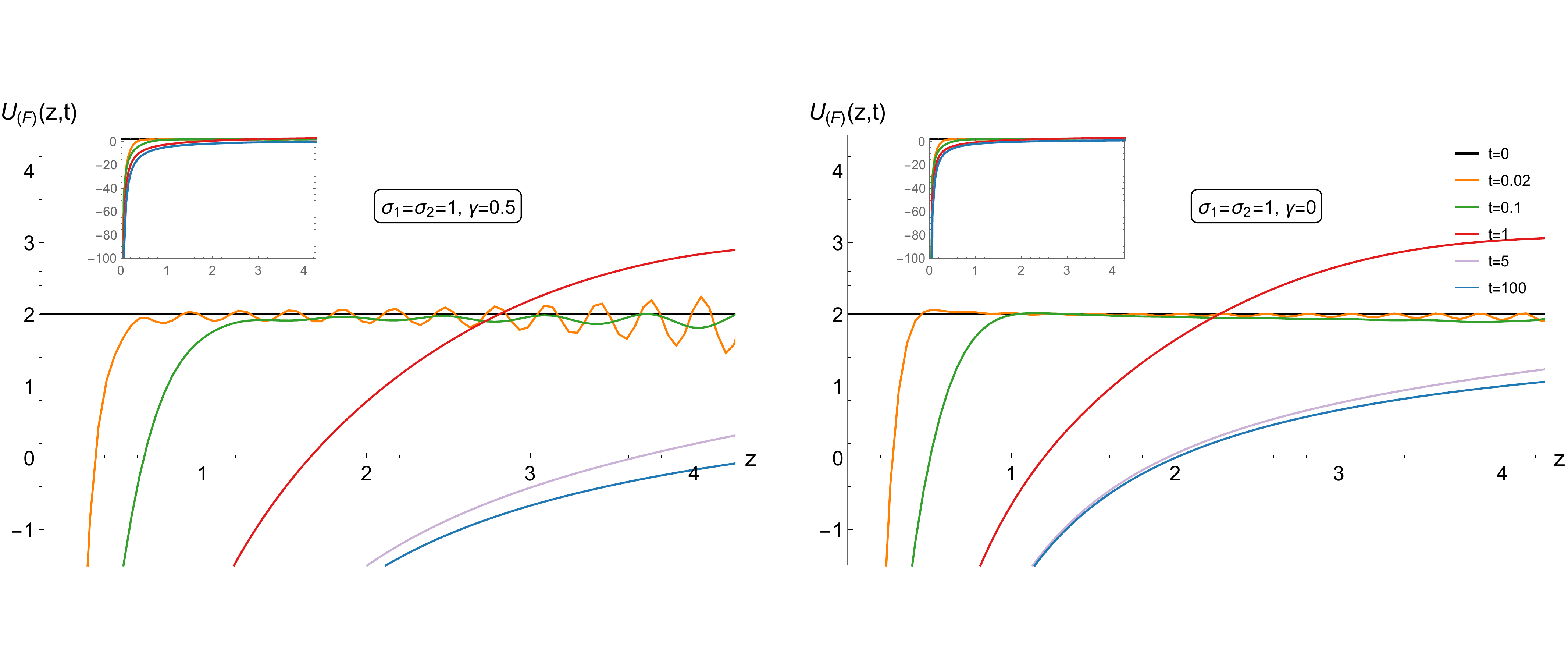}
\caption{Plots for Eq.(\ref{force-burg}) with initial condition $\psi(z)=2$ (shown in black), for parameter values $\sigma_1 = \sigma_2 = 1$, at different $t$ values. The sink parameter $\gamma$ for the left panels and right panels is given as $0.5$ and $0$ respectively. The insets for each panel have the same parameter values and offer a larger range to give more perspective.}
\label{fig:forced-burgers}
\end{center}
\end{figure}

We give example plots of Eq.(\ref{force-burg}) in Figure \ref{fig:forced-burgers}, where the left and right panels match the parameter values $\{\sigma_1, \sigma_2, \gamma \}$ given in Figure \ref{fig:linearforced}. Once again, we used \texttt{NIntegrate} of \textit{Mathematica}\textsuperscript{\textregistered} 11.1.1.0 to evaluate the necessary integrals. Due to the difficulty of automating the process of cutoff-finding for the required two-dimensional integrals involving $\mu$ and $\xi$, and the fact that the corresponding confluent hypergeomtric terms did not present numerical instability as discussed in Section \ref{numcomp}, the integration cutoffs were manually determined for each $z$-value in this case. Additionally, the discrete eigenvalue summations were approximated by computing the first $100$ terms, with their rate of convergence proving sufficient to make the associated error negligible.

In Figure \ref{fig:forced-burgers} we notice very different dynamic behaviour from that witnessed Figure \ref{fig:hierarchy-burgers}.  Most notably, the steady-state behaviour is no longer a constant line at $U=0$, instead given by:
\begin{eqnarray}
\begin{split}
\lim_{t \rightarrow \infty} \frac{\frac{\partial}{\partial z} \int^{\infty}_{0} d\xi G_{(F)} (z,\tau(t)|\xi) \exp \left[ -\frac{\alpha(0)}{2} \int^{\xi} d \chi \psi(\chi) \right]}{ \int^{\infty}_{0} d\xi G_{(F)}(z,\tau(t)|\xi) \exp \left[ -\frac{\alpha(0)}{2} \int^{\xi} d \chi \psi(\chi) \right]} = \frac{\frac{\partial}{\partial z}\sqrt{W(z)} \vartheta_0 (z)}{\sqrt{W(z)} \vartheta_0 (z)},\\
= \underbrace{\frac{1 + 2 \gamma + \sigma_1}{2 z}}_{\textrm{pole at origin}} - \underbrace{\frac{\sigma_2}{2}\left( \frac{1+\gamma+\sigma_1}{1 + 2\gamma + \sigma_1} \right)}_{\textrm{constant term}},\\
\Rightarrow U_{(F)}(z,T\gg 0) = \frac{2}{\alpha(T)}\left[  \frac{\sigma_2}{2}\left( \frac{1+\gamma+\sigma_1}{1 + 2\gamma + \sigma_1} \right) - \frac{1 + 2 \gamma + \sigma_1}{2 z} \right],
\end{split}
\label{forcedstst}
\end{eqnarray}
which is bound between the two (large time) extreme values of $\alpha(T)$
\begin{equation}
\frac{1}{4} \le \alpha(T) \le \frac{3}{4}.
\end{equation}
Focusing on the left plot, we see that for small time ($t=0.02$) the solution profile begins to oscillate around the initial condition of $U_{(F)}=2$ for $z>0.5$. For $z<0.5$, the wave-density profile negatively diverges to a pole at the origin, reflecting the eventual large time behaviour indicated in Eq.(\ref{forcedstst})---a negative pole at the origin and a horizontal asymptote to a constant as $z \rightarrow \infty$. As time progresses, the oscillatory behaviour subsides, and by $t \ge 1$ the profile becomes what we would expect as generated from Eq.(\ref{forcedstst}). Focusing on the inset of the left panel in Figure \ref{fig:forced-burgers}, which zooms out the profiles offered in the parent left panel, we are able to appreciate the negative pole at the origin that forms for $t>0$. The right panel, which differs from the corresponding left panel by having zero sink parameter ($\gamma =0$), offers similar profiles, with the main difference being the oscillations around the initial condition of the $t<1$ plots being markedly diminished in amplitude. Indeed, it is noteworthy that \textit{turning on} the sink parameter in the linear system exaggerates interesting oscillatory behaviour in the corresponding non linear system. Furthermore, we notice the remaining $t \ge 1$ profiles show greater values for $z \ge 4$, which is a consequence of the constant term in Eq.(\ref{forcedstst}) obtaining a higher value for $\gamma = 0$. The corresponding right inset in Figure \ref{fig:forced-burgers}, which again zooms out the profiles offered in the parent right panel, enables us to appreciate the negative pole at the origin that forms for $t>0$.

\section{Conclusions and future work}
\label{concandfut}
In this work we have provided the means to solve the general Cauchy problem for the $N$-th order equation in the Burgers hierarchy, and the forced Burgers equation whose corresponding linearised equation is the sinked Bessel process with constant drift. Additionally, both systems have general time dependent coefficients in their defining differential equations, meaning that they may not be amenable to steady-state analysis. 

The Green's function solution associated with the forced Burgers equation presented in this work---the sinked variant of the Bessel process with constant drift---is a genuinely new result. Using the analytic form of this Green's function we were able to expose a potentially counter-intuitive result: that the parameter which controls sinking in the linear system, also controls the amplitude of interesting oscillatory behaviour in the non linear system. Furthermore, exploiting the main result of \cite{GORSKA13}, we have provided analytic solutions to the Green's functions associated with the Burgers hierarchy---Meijer-G/generalised hypergeometric function solutions to $N$-th order HOHTEs. Numerical solution of the associated Cauchy problem is challenging as the inherent complexity of the Green's function leads to numerical instabilities, especially for higher-$N$. These challenges were overcome by pre-processing the integrands in order to obtain acceptably negligible integral error terms, while simultaneously avoiding numerical instability for the hypergeometric terms.

From an analytic perspective, the results in this work may offer insights into other NLPDEs, both in regards to their Cauchy problem for general initial conditions, and other methods of finding exact solutions through techniques such as the G'/G-expansion method \cite{HE13}. Furthermore, the methods of solution detailed in this work may readily be used to provide relatively simple validation of numerical integration algorithms. This offers a means to test NLPDEs of any order $N$ and initial condition. 

\section*{Acknowledgements}
The authors would like to thank Alexander Kalloniatis for providing invaluable feedback to an early version of this manuscript. We additionally thank the anonymous reviewers for pointing out errors during the review process.

\appendix
\section{Derivation of continuous spectrum normalisation: MacRobert's proof} \label{Appendix1}

We compute the normalisation constant $\rho(\mu)$ by evaluating the inverse integral transform in Eq.(\ref{MacRob}), given explicitly as
\begin{eqnarray}
\begin{split}
{\cal I}(\nu,y) = \int^{\infty}_0 dx W(x) \vartheta_{a,b}(\nu,x) \int^{\kappa_2}_{\kappa_1} d\mu \rho(\mu) \vartheta_{a,b}(\mu,x) \vartheta_{a,b}(\mu,y)
\label{MacRob1}
\end{split}
\end{eqnarray}
where $\{\nu, y, \kappa_1,\kappa_2\} \in \mathbb{R}_+$ and $\kappa_1 < \kappa_2$.

In order to proceed we are required to reverse the order of the integrals, refer to Chapter 14 of Davies \cite{Davies02} for similar examples involving Bessel/Hankel functions. To accomplish this we need to bound the behaviour of ${\cal I}(\nu,y)$ in the $x \rightarrow \infty$ limit. As the confluent hypergeometric function has an irregular singularity at this point, we apply the confluent hypergeometric connection formula (refer to Chapter 7 of Olver \cite{Olver74}) 
\begin{equation}
\vartheta_{a,b}(\mu,x) = e^{-\frac{\pi a}{\mu}} \left[ \frac{e^{i \pi b}}{\Gamma\left( b- \frac{i a}{\mu} \right)} \chi_{a,b}(\mu,x) +  \frac{e^{-i \pi b}}{\Gamma\left( b+ \frac{i a}{\mu} \right)} \chi^*_{a,b}(\mu,x) \right],
\label{connection}
\end{equation}
where 
\begin{equation}
\chi_{a,b}( \mu,x) =x^{\gamma}e^{\left(\frac{\sigma_2}{2} - i\mu \right)x} \Psi \left(\left. b + \frac{i a}{\mu}, 2b \right|  i2 \mu x \right),
\label{tricomi}
\end{equation}
and $\Psi$ is the Tricomi confluent hypergeometric function which has the following asymptotic properties,
\begin{equation}
\lim_{x \rightarrow \infty} \Psi\left(\left. b + \frac{i a}{\mu}, 2b \right|  i2 \mu x \right) \sim (  i 2 \mu x )^{-b - \frac{i a }{\mu}}\,_2F_0\left(  \left. b + \frac{i a}{\mu}, 1-b  + \frac{i a}{\mu}  \right|  \frac{ i}{  2 \mu x} \right).
\end{equation}
\begin{figure}
\begin{center}
\includegraphics[height=4cm]{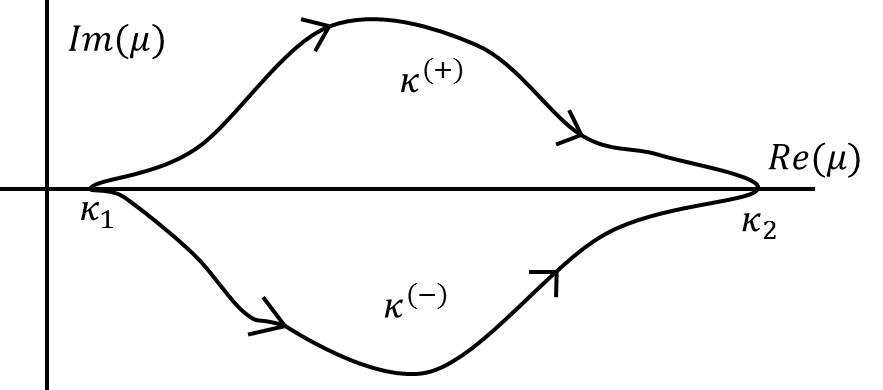}
\caption{Deformed contours $\kappa^{(\pm)}$ in the $\mu$-plane}
\label{fig:MacRob}
\end{center}
\end{figure}
Hence, applying Eq.(\ref{connection}) to Eq.(\ref{MacRob1}), and deforming the $\mu$ integrals onto the contours $\kappa^{(+)}$ and $\kappa^{(-)}$ as shown in Figure (\ref{fig:MacRob}) we obtain
\begin{eqnarray}
\begin{split}
{\cal I}(\nu,y) = \int^{\infty}_0dx W(x) \vartheta_{a,b}(\nu,x)\int_{\kappa^{(-)}}d\mu \rho(\mu) \frac{e^{-\pi \left( \frac{a}{\mu} - ib \right)}}{\Gamma \left( b - \frac{i a}{\mu} \right)} \chi_{a,b} (\mu,x)\vartheta_{a,b}(\mu,y)\\
+  \int^{\infty}_0dx W(x) \vartheta_{a,b}(\nu,x)\int_{\kappa^{(+)}}d\mu \rho(\mu) \frac{e^{-\pi \left( \frac{a}{\mu} + ib \right)}}{\Gamma \left( b + \frac{i a}{\mu} \right)} \chi^*_{a,b} (\mu,x)\vartheta_{a,b}(\mu,y).
\end{split}
\end{eqnarray}
In this form we may reverse the order of integration as the particular choice of contour $\kappa^{(\pm)}$ ensures that each term is exponentially damped by the term $e^{-|Im(\mu)|x}$ as $x \rightarrow \infty$. Hence, ${\cal I}(\nu,y)$ becomes
\begin{eqnarray}
\begin{split}
{\cal I}(\nu,y) = \int_{\kappa^{(-)}}d\mu \rho(\mu) \frac{e^{-\pi \left( \frac{a}{\mu} - ib \right)}}{\Gamma \left( b - \frac{i a}{\mu} \right)}\vartheta_{a,b}(\mu,y) \int^{\infty}_0dx W(x) \vartheta_{a,b}(\nu,x) \chi_{a,b} (\mu,x)\\
+ \int_{\kappa^{(+)}}d\mu \rho(\mu) \frac{e^{-\pi \left( \frac{a}{\mu} + ib \right)}}{\Gamma \left( b + \frac{i a}{\mu} \right)}\vartheta_{a,b}(\mu,y) \int^{\infty}_0dx W(x) \vartheta_{a,b}(\nu,x) \chi^*_{a,b} (\mu,x).
\label{contour}
\end{split}
\end{eqnarray}
Noting that $\{ \vartheta_{a,b}(\mu,x), \chi_{a,b} (\mu,x) \}$ are eigenfunctions of the same (real) governing Sturm-Liouville operator
\begin{equation}
\left[{\cal H} +\left(  \frac{\sigma^2_2}{4}+\mu^2 \right)\right] \{  \vartheta_{a,b}(\mu,x), \chi_{a,b} (\mu,x)\} = 0,
\label{S-L2}
\end{equation}
for ${\cal H}$ given in Eq.(\ref{S-L}), we perform the following transformation to the eigenfunctions
\begin{equation}
\sqrt{W(x)}\{  \vartheta_{a,b}(\mu,x), \chi_{a,b} ( \mu,x)\} = \left\{  \tilde{  \vartheta}_{a,b}(\mu,x), \tilde{\chi}_{a,b} (\mu,x) \right\},
\end{equation}
which transforms Eq.(\ref{S-L2}) into
\begin{equation}
\left[ \frac{d^2}{dx^2}+\left( \mu^2 -\frac{2a}{x} + \frac{b(b-1)}{x^2} \right) \right]\left\{  \tilde{  \vartheta}_{a,b}(\mu,x), \tilde{\chi}_{a,b} (\mu,x) \right\}=0,
\label{S-L3}
\end{equation}
which is the form required to evaluate the $x$-integrals in Eq(\ref{contour}). Note that the reality of the operators in Eq.(\ref{S-L2}) and (\ref{S-L3}) also ensure that the same arguments also hold for $\tilde{\chi}^*_{a,b} (\mu,x)$. 

In order to accomplish our task, we consider two copies of Eq.(\ref{S-L3}) for both $ \tilde{  \vartheta}_{a,b}(\nu,x)$ and $\tilde{\chi}_{a,b} (\mu,x) $. Multiplying the equation for $\tilde{  \vartheta}_{a,b}(\nu,x)$ by $\tilde{\chi}_{a,b} (\mu,x)$, and the equation for $\tilde{\chi}_{a,b} ( \mu,x)$ by $\tilde{  \vartheta}_{a,b}(\nu,x)$, and subtracting the former from the latter we obtain,
\begin{equation}
\tilde{  \vartheta}_{a,b}(\nu,x) \tilde{\chi}_{a,b} ( \mu,x) =  \frac{\tilde{\chi}_{a,b} (\mu,x) \frac{d^2}{dx^2}\tilde{  \vartheta}_{a,b}(\nu,x) -  \tilde{  \vartheta}_{a,b}(\nu,x) \frac{d^2}{dx^2}\tilde{\chi}_{a,b} (\mu,x) }{\mu^2-\nu^2},
\label{S-L4}
\end{equation}
and similarly for $\tilde{  \vartheta}_{a,b}(\nu,x) \tilde{\chi}^*_{a,b} ( \mu,x)$. Integrating Eq.(\ref{S-L4}) over $x \in \mathbb{R}_+$ and applying integration by parts results in
\begin{eqnarray}
\begin{split}
\int^{\infty}_0 dx \tilde{  \vartheta}_{a,b}(\nu,x) \tilde{\chi}_{a,b} (\mu,x) =  \int^{\infty}_0 dx W(x) \vartheta_{a,b}(\nu,x) \chi_{a,b} (\mu,x),\\
= \left[ \frac{\tilde{\chi}_{a,b} (\mu,x) \frac{d}{dx}\tilde{  \vartheta}_{a,b}(\nu,x) -  \tilde{  \vartheta}_{a,b}(\nu,x) \frac{d}{dx}\tilde{\chi}_{a,b} (\mu,x) }{\mu^2-\nu^2} \right]^{x \rightarrow \infty}_{x \rightarrow 0}.
\end{split}
\label{limit1}
\end{eqnarray}
Substituting this expression, and the corresponding expression involving $\tilde{\chi}^*_{a,b} (\mu,x)$  into Eq.(\ref{contour}) for ${\cal I}(\nu,y)$ transforms the integral over $x$ into a more manageable series of limits. Moreover, by application of the \textit{inverse} confluent hypergeometric connection formula found in Chapter 7 of \cite{Olver74}
\begin{equation}
\tilde{\chi}_{a,b} (\mu,x) = \frac{\pi}{\sin(2\pi b)}\left[ \frac{\tilde{  \vartheta}_{a,b}(\mu,x)}{\Gamma\left( 1-b+ \frac{i a}{\mu} \right)} -( i 2 \mu )^{1-2b} \frac{\tilde{  \vartheta}_{a,1-b}(\mu,x)}{\Gamma\left( b+ \frac{i a}{\mu} \right)}\right],
\end{equation}
we can immediately determine that Eq.(\ref{limit1}) in the limit $x \rightarrow 0$ is 0.  

In determining the value of the remaining limit to compute ${\cal I}(\nu,y)$, we apply the connection formula (Eq.(\ref{connection})) to the eigenfunctions with eigenvalue $\nu$ in Eq.(\ref{limit1}). Then using the following asymptotic expressions
\begin{eqnarray}
\begin{split}
\lim_{x \rightarrow \infty}\tilde{\chi}_{a,b}(\mu, x ) \sim ( 2 \mu)^{-b- \frac{i a}{\mu}} x^{- \frac{i a}{\mu}} \exp \left[ \frac{\pi a}{2 \mu} - i \left( \frac{\pi b}{2} + \mu x \right) \right]  \left( 1+O\left( x^{-1} \right) \right),\\
\lim_{x \rightarrow \infty}\frac{d}{dx}\tilde{\chi}_{a,b}( \mu, x ) \sim -i \mu( 2 \mu)^{-b- \frac{i a}{\mu}} x^{- \frac{i a}{\mu}} \exp \left[ \frac{\pi a}{2 \mu} - i \left( \frac{\pi b}{2} + \mu x \right) \right]  \left( 1+O\left( x^{-1} \right) \right),
\end{split}
\end{eqnarray}
${\cal I}(\nu,y)$ becomes
\begin{eqnarray}
\begin{split}
{\cal I}(\nu,y) = \lim_{\varkappa \rightarrow \infty} \left\{  \int_{\kappa^{(-)}}d\mu \frac{\rho(\mu)e^{-\frac{\pi a}{2}\left( \frac{1}{\mu}+\frac{1}{\nu} \right)}\vartheta_{a,b}(\mu,y) }{(4 \mu \nu)^b}  \left[ {\cal K}_1(\mu,\nu) \left(\frac{i e^{-i(\mu+\nu)\varkappa} }{\mu+\nu} \right)  \right. \right.\\
 \left. + {\cal K}_2(\mu,\nu) \left(\frac{i e^{-i(\mu-\nu)\varkappa} }{\mu-\nu} \right) \right] 
- \left.  \int_{\kappa^{(+)}}d\mu  \frac{\rho(\mu)e^{-\frac{\pi a}{2}\left( \frac{1}{\mu}+\frac{1}{\nu} \right)}\vartheta_{a,b}(\mu,y) }{(4 \mu \nu)^b}  \right. \\
 \left.  \left[ {\cal K}^*_1(\mu,\nu) \left(\frac{i e^{i(\mu+\nu)\varkappa} }{\mu+\nu} \right)+ {\cal K}^*_2(\mu,\nu) \left(\frac{i e^{i(\mu-\nu)\varkappa} }{\mu-\nu} \right) \right] \right\},
\end{split}
\label{contour2}
\end{eqnarray}
where we have labeled $\varkappa = x + a (\mu \nu)^{-1} \log_e (x)$, and
\begin{equation}
{\cal K}_1(\mu,\nu) =  \frac{  (2 \mu)^{ - i \frac{a}{\mu}} (2 \nu)^{- i \frac{a}{\nu}}e^{ i \pi b}}{\Gamma \left( b-i\frac{a}{\mu} \right)\Gamma \left( b-i\frac{a}{\nu} \right)}, \,\, 
{\cal K}_2(\mu,\nu) =  \frac{  (2 \mu)^{- i \frac{a}{\mu}} (2 \nu)^{ i \frac{a}{\nu}}}{\Gamma \left( b-i\frac{a}{\mu} \right)\Gamma \left( b+i\frac{a}{\nu} \right)}.
\end{equation}
Expressing the exponential terms involving $\varkappa$ in Eq.(\ref{contour2}) as the corresponding trigonometric functions using Euler's identity, we obtain 
\begin{eqnarray}
\begin{split}
{\cal I}(\nu,y) = \lim_{\varkappa \rightarrow \infty} 2 \mathbb{R} \int^{\kappa_2}_{\kappa_1}d\mu \frac{\rho(\mu)e^{-\frac{\pi a}{2}\left( \frac{1}{\mu}+\frac{1}{\nu} \right)} }{(4 \mu \nu)^b} \vartheta_{a,b}(\mu,y) \left[{\cal K}_1(\mu,\nu)  \left( \frac{\sin (\mu + \nu)\varkappa}{\mu + \nu} \right. \right. \\
\left.+ \frac{ i \cos (\mu + \nu) \varkappa}{\mu + \nu} \right)  
+ \left. {\cal K}_2(\mu,\nu) \left( \frac{\sin (\mu - \nu)\varkappa + i \cos (\mu - \nu) \varkappa}{\mu - \nu} \right) \right],
\end{split}
\label{contour3}
\end{eqnarray}
where we have deformed the contours $\kappa^{(-)}$ and $\kappa^{(+)}$ in the $\mu$-plane back to the positive real line segment $(\kappa_1,\kappa_2)$ as originally presented in Eq.(\ref{MacRob1}). The particular combination of limit and integral involving trigonometric functions in Eq.(\ref{contour3}) is commonly referred to as a \textit{Dirichlet integral} (refer to Chapter 1 of MacRobert \cite{MACROBERT}). Importantly, for the analytic function $M(\xi)$ which possesses only a finite number of discontinuities and turning points on the interval $(-\kappa,\kappa)$, $\kappa \in \mathbb{R}$, the following identities hold
\begin{equation}
\lim_{\varkappa \rightarrow \infty}\int^{\kappa}_{-\kappa} d\xi M(\xi) \frac{\cos (\xi \varkappa)}{\xi} = 0, \,\, \lim_{\varkappa \rightarrow \infty}\int^{\kappa}_{-\kappa} d\xi M(\xi) \frac{\sin (\xi \varkappa)}{\xi} = \frac{\pi}{2}[ M(0+)+ M (0-) ].
\label{dirich}
\end{equation}
Thus, if we assume that $\rho(\mu)$ is a suitably well behaved function which enables Eq.(\ref{dirich}) to apply, and taking the limits $\kappa_1 \rightarrow 0$ and $\kappa_2 \rightarrow \infty$ in Eq.(\ref{contour3}), we finally obtain from Eq.(\ref{MacRob})
\begin{eqnarray}
\begin{split}
\vartheta_{a,b}(\nu,y) = \frac{2 \pi  \rho(\nu) e^{-\frac{\pi a }{\nu} }  \vartheta_{a,b}(\nu,y) {\cal K}_2 (\nu,\nu) }{(2 \nu)^{2b}}\\
 \Rightarrow \rho(\nu) = \frac{(2 \nu)^{2b} e^{\frac{\pi a}{\nu}} \left| \Gamma\left( b+ i \frac{a}{\nu} \right) \right|^2}{2 \pi}.
\end{split}
\end{eqnarray}

\end{document}